# How the Massachusetts Assault Weapons Ban Enforcement Notice Changed Firearm Sales

February 2021

Meenakshi Balakrishna and Kenneth C. Wilbur[*]
University of California, San Diego

The Massachusetts Attorney General issued an Enforcement Notice in 2016 to announce a new interpretation of a key phrase in the state's assault weapons ban. The Enforcement Notice increased sales of tagged assault rifles by 616% in the first 5 days, followed by a 9% decrease over the next three weeks. Sales of Handguns and Shotguns did not change significantly. Tagged assault rifle sales fell 28-30% in 2017 compared to previous years, suggesting that the Enforcement Notice reduced assault weapon sales but also that many banned weapons continued to be sold. Tagged assault rifles sold most in 2017 in zip codes with higher household incomes and proportions of white males. Overall, the results suggest that the firearm market reacts rapidly to policy changes and partially complies with firearm restrictions.

Acknowledgements and disclaimer:
We gratefully acknowledge the Massachusetts Department of Criminal Justice Information Services for providing data; Arnab Pal for first identifying the data source; Snehanshu Tiwari, Ganesh Chandrasekaran Iyer, Jie Chen, Immanuel Kwok, and Rebecca Wang for superlative research assistance; and John Donohue, Jessica Kim, Kanishka Misra, Yi-Lin Tsai and Robert Sanders for helpful comments and suggestions. Any remaining errors are ours alone.

Corresponding author: kcwilbur@ucsd.edu

**I. INTRODUCTION**

The *Mass Shooting Tracker* database recorded 427 mass shootings in America in 2017,[1] resulting in 590 deaths, 4.1% of U.S. firearm homicides.[2] Mass shootings that involve semiautomatic weapons produce more injuries and deaths (de Jager et al. 2018) and have sometimes precipitated changes in firearm policy (Luca, Poliquin and Malhotra 2019), some of which have reduced mass shootings (Donohue and Boulouta 2019; Lemieux, Prenzler and Bricknell 2013).[3] Military-style semiautomatic firearms are banned in Australia, Canada, the European Union, New Zealand, the U.K. and 7 U.S. states. Walmart stopped selling assault-style rifles in 2015, followed by Dick's Sporting Goods in 2018. 67% of U.S. adults surveyed said that they favor banning assault-style weapons (Pew Research 2017).

The U.S. federal government banned "semiautomatic assault weapons" from 1994 until 2004, defining them as the following categories of firearms:[4]

- 9 specific firearm products (e.g., Colt AR-15, AK-47, etc.), including "copies and duplicates" of those specific products.
- Semiautomatic rifles that accept detachable magazines and offer at least two of five features: folding/telescoping stock, pistol grip, bayonet mount, flash suppressor, and grenade launcher.
- Semiautomatic pistols and semiautomatic shotguns with at least two features among similar feature sets.

---

[1] The database defines a mass shooting as an incident in which four or more people are shot.
[2] https://www.cdc.gov/nchs/data/nvsr/nvsr68/nvsr68_09-508.pdf, Table 8.
[3] The U.S. federal government banned new machine guns in 1986, importations of semiautomatic rifles in 1989 and 1998, sales of new semiautomatic assault weapons from 1994 until 2004, and bump stocks in 2018.
[4] H.R. 3355-203, Title XI, sec. 110102. https://www.congress.gov/bill/103rd-congress/house-bill/3355/text

The definition reflected a political compromise (Lenett 1995). Several of the specific firearm products, most notably the Colt AR-15, were already off patent and readily extensible when the ban was passed in 1994. After the statute was enacted, firearm manufacturers quickly introduced minor variations on the specific firearm products named in the ban, thereby skirting the "copies and duplicates" restriction. Many variants were based on the Colt AR-15, the semiautomatic version of the fully automatic M16 machine gun. Derivatives of the AR-15 are called "AR-15-style," "AR-style," or "Modern Sporting Rifles." An estimated 1-2 million AR-style rifles were manufactured in 2016[5] with 8.5-15 million in circulation.[6] Colt retired its consumer-market AR-15 in 2019, stating that "the market for modern sporting rifles has experienced significant excess manufacturing capacity."[7]

The purpose of the current paper is to estimate the causal effects of a natural experiment in firearm policy on firearm sales. On the morning of July 20, 2016, Massachusetts Attorney General Maura Healey publicly announced an Enforcement Notice ("EN") to provide the first legal interpretation of the key phrase "copies and duplicates" in the definition of banned assault weapons. The EN was announced in a press conference and a *Boston Herald* editorial and generated substantial immediate publicity in the state. The AGO also notified the state's firearms retailers of the change in letters it mailed in the afternoon of July 18, 2016. The announcements were not preceded by public comment or debate and were widely seen as a surprise.

Massachusetts requires all firearm dealers to register every new firearm sale with the state. Figure 1 graphs the population of legally registered firearm sales, by type of firearm, for a 6-week period centered on the date of the Enforcement Notice. Two features stand out. First,

---

[5] https://web.archive.org/web/20190821183313/https://www.washingtonpost.com/news/business/wp/2018/02/23/u-s-gun-manufactures-have-produced-150-million-guns-since-1986/
[6] https://www.mcclatchydc.com/news/nation-world/national/article201882739.html
[7] https://www.nytimes.com/2019/09/19/business/colt-ar-15.html

Rifle sales rose more than 20-fold on the date of the EN, without comparable increases in Handgun or Shotgun sales. Second, the series of Rifle sales after the peak appears to be only slightly lower than before the Enforcement Notice. Of course, the graph does not show what sales would have been observed in the absence of the EN; simple pre/post comparisons can be muddled by multiple factors such as seasonality.

We estimate causal effects of the EN on firearm sales. We focus on legal firearm sales as an outcome variable because the express purpose of the assault weapon ban and EN are to restrict acquisitions of new assault weapons. First, we use an automated image classifier to distinguish Rifle sales into two categories: Tagged Assault Weapons ("TAW Rifles") and all non-tagged Rifles ("Non-TAW Rifles"). The classifier is conservative but reasonably accurate, as discussed further below. If the market complied perfectly with the EN, and if the classifier were perfect, then TAW Rifle sales should be zero after the EN announcement. Second, we estimate auto-regressive models to predict counterfactual sales, i.e., what sales would have occurred in the absence of the EN, based on lagged sales, time effects and firearm license issuances by date. Firearm license issuances provide a good instrument for short-term firearm sales, because a license is required to legally purchase or possess a firearm in Massachusetts, and because license applications and renewals normally take 35-40 days to process. The number of licenses issued in the first few weeks after the EN depended solely on the number of license applications filed before the EN, and therefore cannot depend on the EN itself. Third, we compare observations of post-EN firearm sales to the model's predictions in order to estimate immediate and short-term causal effects. Finally, we present a descriptive analysis of longer-term trends in firearm sales after the EN.

We find that TAW Rifle and Non-TAW Rifle sales both rose sharply in the first few days after the EN. In the following three weeks, TAW Rifle sales fell by 9% and Non-TAW Rifle sales fell by 15%, compared to the model's counterfactual predictions, but those changes were not statistically significant. The EN did not significantly change Handgun or Shotgun sales in the short term. Longer term, TAW Rifle sales in 2017 were 28-30% lower than in comparable previous years, whereas sales of other firearm types did not show similar declines. We also investigate where 2017 TAW Rifle sales remained highest, finding that they fell least in retailers located in zip codes with high household incomes and white male populations. The longer-term analysis suggests partial effectiveness of the EN: it seems to have reduced assault weapon sales substantially, but it also shows that many banned assault weapons continued to be sold.

In the next section, we explain the empirical context of the natural experiment and the details of the policy change. We then describe the relevant literature, data, methods and causal evidence, and longer-term data descriptives. We conclude with implications for policy makers and the limitations of the exercise.

**II. EMPIRICAL CONTEXT: THE NATURAL EXPERIMENT**

Massachusetts banned assault weapons in 1998 using the same definition as the 1994 federal assault weapons ban.[8] Unlike the federal law, the Massachusetts ban did not expire, and therefore has operated continuously since its enactment. First-time offenders face fines of $1,000-$10,000 and imprisonment of 1-10 years.[9]

---

[8] https://www.nraila.org/articles/20160722/massachusetts-attorney-general-unilaterally-bans-thousands-of-previously-legal-guns
[9] https://malegislature.gov/Laws/GeneralLaws/PartI/TitleXX/Chapter140/Section131M

On July 20, 2016, the Massachusetts Office of the Attorney General ("AGO") issued a public Enforcement Notice regarding prohibited assault weapons.[10] The purpose was to explain what the office considered to be "copies or duplicates" of the specific firearms banned under the state's legal definition of assault weapons. The phrase "copies or duplicates" was not defined in the 1994 federal statute or the 1998 state law and had not been interpreted by the courts. The EN introduced two new tests that indicate whether a firearm is a "copy or duplicate" of a banned weapon: a *Similarity Test* indicating whether its internal functional components are substantially similar to a banned firearm, and an *Interchangeability Test* indicating whether its receiver is the same as or interchangeable with the receiver of a banned firearm. Because the Colt AR-15 was explicitly banned by the 1998 state law, the Enforcement Notice thereby indicated that the Massachusetts assault weapons ban covered a large number of semiautomatic AR-style rifles that were available for sale in the state. The Enforcement Notice was followed by public debate about whether the AGO had exceeded its legal authority. The EN has since been affirmed by multiple judicial decisions with additional litigation ongoing.[11]

The Enforcement Notice was issued without prior public debate or comment.[12] It came 39 days after a mass shooting in which 53 people died in Orlando on June 12, 2016. Highly publicized mass shootings are known to stimulate immediate legal firearm acquisitions (Studdert et al. 2017).

Contemporaneous AGO communications indicated that the AGO (i) would immediately apply the Enforcement Notice to new firearm sales; (ii) would not enforce the EN retroactively to firearms sold prior to that date, and (iii) did not plan to enforce the EN on relevant weapon transactions that had been initiated prior to July 20, 2016 and not yet been completed.[13] Because the EN applied only to new weapon sales, it did not affect owners' existing assault weapons and would not prevent legal transfers or sales of existing used assault weapons.

---

[10] https://www.mass.gov/files/documents/2018/11/13/assault-weapons-enforcement-notice.pdf
[11] https://www.reuters.com/article/us-usa-guns-massachusetts/u-s-judge-upholds-massachusetts-assault-weapons-ban-idUSKCN1HD2CW, https://www.telegram.com/news/20190728/ag-says-controversial-enforcement-notice-doesnt-apply-to-types-of-rifles-gun-shops-sued-over
[12] https://www.wwlp.com/news/several-types-of-guns-now-banned-in-massachusetts/, https://patch.com/massachusetts/woburn/buyers-crowd-woburn-gun-store-after-ags-enforcement-notice

The AGO did not provide a list of firearm products that were newly banned under the Enforcement Notice. AGO communications accompanying the EN said "The Attorney General expects voluntary compliance from gun dealers and manufacturers with respect to prohibited weapons," and that, "By issuing the notice, the Attorney General hopes and expects that non-compliant gun dealers will come into voluntary compliance with the law, to minimize the need for criminal or civil enforcement."[14]

In 2018, we asked a state government official how to determine whether a specific firearm meets the Similarity Test or the Interchangeability Test. They indicated that such determinations require a physical inspection of the specific weapon.

The EN and related publicity caused a dramatic and immediate reaction in firearm sales, as shown in Figure 1. Given the unanticipated nature of the Enforcement Notice, we assume its timing to be an exogenous shock to the Massachusetts firearm market. We therefore treat it as a "natural experiment" or "quasi-experiment." We use its timing to identify the causal effect of a change in assault weapons policy on firearm sales in the immediate and short run.

The natural experiment is unusual for two reasons. First, a change in the interpretation of a firearm ban occurred without an accompanying change in the statute itself. We are not aware of any other occasion on which the legal definition of a banned firearm changed during the time the ban was continually in effect. Therefore, we can separate the causal effects of the policy change from any related effects of other factors (e.g., legislative debate).

---

[13] https://www.mass.gov/guides/frequently-asked-questions-about-the-assault-weapons-ban-enforcement-notice
[14] https://www.mass.gov/guides/frequently-asked-questions-about-the-assault-weapons-ban-enforcement-notice

Second, although it is straightforward to predict that a broader interpretation of banned assault weapons should reduce assault weapon sales, it is less obvious to predict how large the effect might be. The size of the effect may reveal the extent to which firearm market participants regulate themselves effectively. It is also unclear how sales of non-banned firearms would react. For example, if a broader interpretation of banned assault weapons leads purchasers to substitute other firearms for assault weapons, then sales of other firearms might rise.

Licenses are required to legally purchase or possess a firearm in Massachusetts. They cost $100, are available to both residents and nonresidents, and must be renewed every six years in order to legally acquire or maintain possession of a firearm. License applications take up to 42 days to be processed.[15] State government personnel told us that 35-40 days normally pass between license application and issuance. There is no fast-track or expedited licensing procedure.[16] We use firearm license data to improve predictions of counterfactual firearm sales for several weeks after the EN. The justification for the license instrument is that, because the EN announcement was a surprise, license applications before the EN could not have been driven by the EN. Therefore, license issuances for several weeks after the EN must be unrelated to EN timing.

We limit the time span of the econometric exercise to 25 days after the EN, during the period when firearm licenses should offer a strong instrument. We do not extend the causal-effects analysis beyond that limited threshold because no strong instrument or control group is available for the presidential election season in fall 2016, an event that corresponded to increased firearm sales. However, we do report some longer-term data descriptives which suggest that the EN was partially but imperfectly effective in reducing assault weapon sales on a longer horizon.

---

[15] https://goal.org/ltc-apply/
[16] Massachusetts passed a law phasing out Class B licenses in 2014, but its effect was minor, as Class A licenses and Firearm Identification Cards account for 96% of all licenses issued from 2006-2017, and anywhere from 95-97% of all licenses in every year between 2011 and 2017.

**III. PREVIOUS LITERATURE**

We contribute to a literature that estimates causal effects of firearm policies on social outcomes. Raissian (2016) found that the 1996 Gun Control Act, which prohibited defendants found guilty of domestic violence misdemeanors from possessing firearms, reduced gun-related homicides among female intimate partners and children. Luca, Malhotra, and Poliquin (2017) found that legally-mandated handgun waiting periods reduce gun homicides in the states that pass them. Donohue, Aneja and Weber (2019) found that right-to-carry concealed handgun laws increase overall violent crime.

More narrowly, we study how a change in assault weapons policy impacts firearm markets and compliance. Koper, Woods and Roth (2004) examined how the federal assault weapon ban affected prices and production in the national firearm market, continuing a series of studies (i.e., Roth and Koper 1997, 1999; Koper and Roth 2001, 2002) which Congress mandated and funded as part of the 1994 assault weapons ban. Prices of semiautomatic pistols stayed roughly flat from 1993-1996, whereas prices of substitutable pistols fell significantly over the same time span. Prices of military-style semiautomatic rifles rose 29% from 1993 to 1995, but 1996 prices fell back to 1993 levels. Production of both semiautomatic pistols and semiautomatic rifles increased significantly in 1993 and 1994 before reverting to near baseline levels in 1995 and falling below baseline in 1996.

Koper, Woods and Roth (2004) noted several aspects of their analysis that also describe the current paper. The authors pointed out that their time-series analyses were "largely descriptive, so causal inferences must be made cautiously." They also noted that "our studies of the [assault weapon] ban have shown that the reaction of manufacturers, dealers, and consumers to gun control policies can have substantial effects on demand and supply for affected weapons both before and after a law's implementation. It is important to study these factors because they affect the timing and form of a law's impact on the availability of weapons to criminals and, by extension, the law's impact on gun violence." Such considerations were "largely absent from prior research" on firearm policies.

Another closely related study by Koper (2014) linked new firearm acquisitions to guns used in crimes. The analysis connected 72,000 legal firearm sales in Maryland to a national database of crime guns recovered by police, identifying more than 1,800 firearms present in both data sets. Among other factors, semiautomatic weapons were disproportionately likely to be recovered by police as crime guns. A state regulation of the used firearm market did not appear to change the probability of a gun's recovery, possibly due to unclear enforcement of the statute.

Finally, Studdert et al. (2017) examined the reaction of legal firearm sales in California to two highly publicized mass shootings in Newtown and San Bernardino. It found the two shootings increased handgun sales by 53% and 41%, respectively. The increases were larger among first-time handgun purchasers. The San Bernardino shooting increased handgun acquisitions by 85% within San Bernardino, as opposed to a 35% increase in the rest of the state.

All comprehensive reviews of the firearm literature agree that more research is needed, including research on how policy influences the market for firearms. As IOM and NRC (2013) explained, "Basic information about gun possession, acquisition, and storage is lacking." They further noted that "fundamental questions about the effectiveness of interventions—both social and legal—remain unanswered." Similarly, Smart et al. (2020, Recommendation 9) suggested that, "to improve understanding of outcomes of critical concern to many in gun policy debates, the U.S. government and private research sponsors should support research examining the effects of gun laws on a wider set of outcomes, including … the gun industry."

We also investigate how retailers responded to the Enforcement Notice, detecting some patterns that shed light on sellers' compliance with the EN. This theme relates to "forensic economics," which Broulik (2020) describes as "economics informing any enforcement stage in any field of law," and Zitzewitz (2012) describes by saying "the economist sizes the extent of an activity about which there had been only anecdotal evidence and provides insight into where it is more prevalent and why." Our approach to forensic evidence is based on detecting behavior that is "hidden in plain sight," in the sense that the underlying data are provided directly to state regulatory authorities by firearm retailers. Christie and Schultz (1994) and Christie, Harris and Schultz (1994) used a similar approach to detect collusion by stock market makers. Fisman and Miguel (2007) examined why some United Nations officials in Manhattan left parking tickets unpaid and how officials responded to a change in parking ticket enforcement policies.

To the best of our knowledge, the policy event we study has not received any prior scientific attention. The closest analogue might be other analyses of sudden policy changes, such as Cunningham and Shah (2018), who found that sudden decriminalization of sex work in Rhode Island increased the prostitution market while simultaneously reducing rape reports and gonorrhea incidence.

Because of the unanticipated nature of the event and the availability of a good instrument, we provide rare causal evidence of how firearm policy changed legal firearm sales. We think this evidence might help policy makers, both inside and outside of Massachusetts, understand the consequences of the Massachusetts Enforcement Notice, as similar policy changes could potentially be adopted in other jurisdictions. Second, we believe this paper provides the first evidence of the rapidity of firearm market participants' response to restrictive firearm policies. Third, we provide long-run descriptive evidence that the Enforcement Notice reduced sales of semiautomatic assault weapons, but that many banned weapons continued to be sold. We show ways the data can help to identify noncompliant firearm retailers; such information might be valuable to law enforcement and sheds light on the extent of firearm industry policy complaince. Finally, we introduce and evaluate a noisy approach to distinguish assault weapons from non-assault weapons, which could potentially facilitate further academic research and efforts to detect and deter weapon sales in jurisdictions where assault weapons are illegal or restricted.

**IV. DATA**

*A. Provenance*

Each time a firearm dealer in Massachusetts sells a firearm, state law requires it to record the transaction date, the purchaser, firearm type, make and model and other information into a state-owned database. We obtained anonymous Massachusetts firearm sales and license data by filing a Freedom of Information Act request with the Massachusetts Department of Criminal Justice Information Services.

We focus on the population of 954,015 legally registered dealer sales that occurred between Jan. 1, 2006 and April 1, 2018. 815,052 sales records occurred prior to the EN. Each record indicates the transaction date, transaction type (sale, registration or transfer), Firearm

Make, Firearm Model, Firearm Type (Handgun, Rifle or Shotgun), Dealer Name, Dealer Shop City, an anonymous purchaser identifier and purchaser gender.

Handguns accounted for 61.3% of all sales, followed by Rifles (26.8%) and Shotguns (11.8%).[17] We do not analyze gender as purchasers were overwhelmingly male at 93%. We do not observe transaction prices.

*B. Distinguishing Assault Rifles from Other Rifles*

We seek to distinguish the effect of the EN on banned assault weapons from its effect on other types of firearms. There were 2,573 distinct Rifle make/model combinations listed in the data. We algorithmically performed the following three steps for each Rifle transaction record.

1) Search Google Images for the Rifle make and model.

2) Download the first image returned in the search results.

3) Assess each unique image using the Google Cloud Vision Application Programming Interface, which provides semantic tags and associated confidence levels.

We separated Rifle transactions into two categories. Images that were tagged as "Assault Rifle" or "Machine Gun" with a confidence level of 80% or greater are classified as *Tagged Assault Weapons* ("TAW Rifles").[18] All other Rifles are counted as *Non-Tagged Assault Weapons* ("Non-TAW Rifles"). 33.7% of all Rifle sales are classified as TAW Rifles. Figure 2 provides the product images and classifications for the 6 top-selling Rifle make/model combinations.[19]

---

[17] The state also tracks Machine Gun sales, but they account for only 0.03% of the transactions and are subject to distinct federal restrictions on transfer and ownership, so we exclude them from the analysis.

[18] The original AR-15 rifles are visually indistinguishable from M-16 machine guns. The main difference is whether the firearm is semiautomatic or fully automatic. https://gundigest.com/reviews/the-ar-16m16-the-rifle-that-was-never-supposed-to-be

[19] We considered whether the Google Vision API could distinguish between assault and non-assault handguns, or between assault and non-assault shotguns, but data quality checks did not provide similar levels of confidence.

We sought to evaluate the accuracy of the Google Vision image classifier. We employed three research assistants (RAs) to create an auxiliary dataset for comparison to the Google Vision results. We provided the legal definition of semiautomatic assault rifles and asked RAs to independently develop a full understanding of its component parts (e.g., "semiautomatic," "detachable magazine," "folding or telescoping stock," etc.). We discussed the definitions with them to ensure that they had successfully followed the instructions. We then provided a list of the 98 highest-selling Rifle make/model combinations, which collectively accounted for 50.0% of all Rifle sales observed in the data. We asked the RAs to search online for manufacturer specifications, product listing pages and product reviews for one Rifle at a time; this was a time-intensive task. Then, based on their best understanding, indicate whether each Rifle make/model combination was a semiautomatic assault weapon according to the legal definition; not a semiautomatic assault weapon; or indeterminate based on missing information. The RAs worked independently without a time limit. They recorded and provided online sources of information for each Rifle so we could evaluate the information they used in reaching their judgments. Our manual review of the RAs' information sources suggested that they took the task seriously and performed it competently.

The RAs are a more authoritative source of assault weapon classifications than Google Vision. Whereas Google Vision accessed a single product image for each Rifle, the RAs accessed extensive product information and multiple images. However, some classifications are inherently ambiguous. The most prominent example is the Ruger Mini-14, the fourth-highest selling Rifle make/model combination with a 1.61% market share. Ruger offered many variants of the Mini-14; including the "Mini-14 Ranch" variants which do not meet the two-feature test in the federal definition of assault weapons, but also several "Ruger Mini-14 Tactical" variants, among which the 5888 trim includes three assault weapon features: folding stock, pistol grip and flash suppressor.[20] Whereas most Ruger Mini-14 variants look similar to traditional hunting rifles, the Mini-14 Tactical presents a decidedly military aesthetic. Unfortunately, the data frequently do not distinguish the variant within a Rifle make/model combination. Therefore, the wide variety among the variants of the Ruger Mini-14 makes its make/model classification as an assault weapon inherently ambiguous, because accurate classification requires the oft-missing variant identifiers. In this particular example, two out of the three RAs classified the Ruger Mini-14 as an assault weapon, so the classification depends on whether we require unanimity among the RAs as a measure of truth, or whether we take the simple majority among the RAs' classifications. We report classifier accuracy for both standards.

Figure 3 provides confusion matrices to assess the accuracy of the Google Vision classifier. The first two matrices take the median RA classification as a measure of truth. The first confusion matrix weights each Rifle make/model equally, finding that the Google Vision API has an overall accuracy rate of 85%. It further shows that the Google Vision classifier is conservative, with a false negative rate of 25% and a false positive rate of 6%. Therefore, when the classifier is wrong, it is more likely to incorrectly categorize a Rifle as a Non-TAW weapon rather than incorrectly categorize a Rifle as a TAW Rifle.

The second confusion matrix weights observations according to sales indices, to provide a better indication of how accurate the classifier would be for a typical sales transaction observed in the data. Specifically, each Rifle's weight is set as its own sales divided by the average sales of the 98 most frequently sold Rifles, so the indices average to one. Under this standard, the classifier has a 75% overall accuracy rate. Its assessments remain conservative, with a false negative rate of 42% and a false positive rate of 7%.

---

[20] https://web.archive.org/web/20201112021945/https://ruger.com/products/mini14TacticalRifle/specSheets/5888.html, accessed January 2021.

RA disagreements primarily indicate inconclusive classifications such as the Ruger Mini-14. The final two confusion matrices restrict attention to the 69 Rifles for which all three RAs indicated the same classification. Dropping ambiguous rifles is essentially restricting attention to the easier subset of Rifle classifications. Within this subset, unweighted classifier accuracy is 94%, and the sales-weighted accuracy rate is again 94%. The classifier remains conservative by this standard, with false positive rates of 0%, and false negative rates of 13% (unweighted) and 16% (share-weighted).

On an absolute scale, the classifier's performance seems acceptable given the complexity of determining whether a Rifle meets the federal definition of a semiautomatic assault weapon. Although it is far from perfect, it tends to undercount assault rifles, on both unweighted and sales-weighted bases. This is not unexpected, as the classifier most easily recognizes AR-style rifles, but there are numerous rifles that fit the strict federal definition of semiautomatic assault rifles but are visually dissimilar to traditional AR-style rifles.[21] Our preference is to avoid inflating the counts of TAW Rifle sales, however the downside of accepting the TAW classifier is that the counts of Non-TAW Rifle sales will likely include many incorrectly classified assault rifles. We conclude that the TAW/Non-TAW Rifle distinction is far from perfect, but it appears to be reasonably accurate, meaningful and conservative, and therefore able to offer deeper insights into how the EN affected firearm sales.

---

[21] The alternative would be paying RAs to code all, or most, of the 2,573 distinct Rifle make/model combinations observed in the data. However, this approach would be expensive and yet would remain subject to ambiguity due to missing firearm variant information.

*C. Trends in Firearm Sales and Licenses*

Figure 4 shows firearm sales by type and year. Annual sales grew more than four-fold from 2007 until 2017, from 27,174 to 111,405. The average annual growth rate was 21% over that period, including a 7% decline in 2014 and a 13% decline in 2017. The Massachusetts population grew 5% over the same ten-year period.[22]

Handgun sales grew the most and the fastest. Handgun sales increased by 442% over ten years, from 16,600 in 2007 to 73,479 in 2017. All four firearm types decreased in 2017 compared to 2016, with declines ranging from 41% for TAW Rifles to 9% for Handguns.

TAW Rifle sales exhibit a different pattern from the other firearm types. TAW Rifle sales grew more than 7-fold from 2007 to 2013, from 1,906 to 14,402. However, sales fell by more than 60% after their 2013 peak, down to 5,980 in 2017. It is unclear how much of the sales reduction is attributable to the EN as opposed to other factors such as market saturation.

We also observe data on all Massachusetts firearm license issuances, classified according to whether they are new licenses or license renewals.[23] Figure 5 shows license issuances by new/renewal status and year. A license remains valid for six years, a pattern which can be seen clearly in the license renewals time series. From 2006-2011, there were 97,658 new licenses and 207,784 license renewals issued. From 2012-2017, there were 209,993 new licenses and 260,936 license renewals issued. Thus, the number of legally permitted firearm owners increased by about half, from about 4.6% to 6.9% of the state population, during the latter six-year period.

Figure 6 shows that daily license issuances did not change discontinuously around the time of the EN. They reached a local maximum about one week after the EN, a spike that likely corresponded to new applications filed in the wake of the mass shooting in Orlando on June 13, 2016.

---

[22] https://www.statista.com/statistics/551711/resident-population-in-massachusetts/
[23] Resident Class A and Firearms Identification Card licenses account for 96% of all licenses issued. Both license types enable legal firearm purchases and possession.

*D. Data Validation*

Braga and Hureau (2015) studied related Massachusetts firearm data as part of their study of sources of crime guns recovered by the Boston Police Department. They documented that dealers were unable to locate sales records for 2.2% of trace requests, suggesting imperfections in recordkeeping. Therefore, we sought to validate the completeness of the Massachusetts firearm sales transaction data using an external source.

We compared the Massachusetts records to state/year National Instant Criminal Background (NICS) data reported by the U.S. Federal Bureau of Investigation. NICS data count background checks conducted by federally licensed firearm retailers prior to a firearm purchase, with separate counts for Handguns and Long Guns. Comparing annual Massachusetts sales records to annual FBI NICS data between 2006 and 2017, we find a correlation of 0.97 for Handguns and 0.96 for Long Guns. Therefore, it appears that the Massachusetts data correspond highly to the external FBI data source, possibly building some confidence in the completeness of the Massachusetts sales records.

**V. MODEL AND ESTIMATION**

To estimate a causal effect of the Enforcement Notice on firearm sales, we must predict what counterfactual firearm sales would have occurred in the absence of the EN. We estimate regression models using the pre-EN sales data, from January 1, 2006, until July 19, 2016. We then use the estimated model to predict what firearm sales would have been on each date shortly after the EN, if the EN had not been issued.[24]

The regression models daily firearm sales, for each type of firearm, as a function of variables that could not have reacted to the EN: lagged sales before the EN, time fixed effects and trends, and license issuances, which depended on pre-EN license applications. We specify the following regression model:

$$y_{jt} = \sum_{\tau=1}^{T_y} \alpha_{j\tau} y_{j,t-\tau} + x_t \beta_j + \sum_{\tau=1}^{T_z} \gamma_{j\tau} z_{N,t-\tau} + \sum_{\tau=1}^{T_z} \delta_{j\tau} z_{R,t-\tau} + \varepsilon_{jt} \qquad (1)$$

$y_{jt}$ is the log of the number of firearm sales of type $j$ on date $t$; $x_t$ includes weekday fixed effects, week-of-year fixed effects, a holiday fixed effect, and linear and quadratic time trends; $z_{Nt}$ and $z_{Rt}$ are the log of the number of new licenses and license renewals issued on date $t$; and $\varepsilon_{jt}$ is an error term.[25]

*A. Model Specification Details*

Time effects and the numbers of lags were chosen to minimize mean absolute prediction errors in a 10-fold cross-validation exercise. The set of time effects included in $x_t$ reduced prediction errors in a first-order autoregressive model relative to several alternatives, including date-of-year fixed effects, as shown in Table 1.

We chose to include $T_y = 28$ lags of sales by successively adding lags to the model until the first time an additional lag increased the mean absolute prediction error in the validation folds. Then, while including 28 lags of sales, we successively added lags of license variables, finding that $T_z = 10$ lags of license issuances minimized mean absolute prediction error in cross-validation exercises. The qualitative results were relatively insensitive to choices of $T_y$ and $T_z$.

---

[24] Our empirical approach is conceptually similar to the "interrupted time series" strategy that Gonzalez-Navarro (2013) used to study Lojack introduction on car thefts, except that it uses a single geographic unit rather than a panel of geographies, and there is no available analogue to Gonzalez-Navarro's exposure variable (i.e., the number of cars treated with Lojack that could have been stolen).
[25] We added 0.1 to firearm sales and license variables before taking logs to avoid taking the log of zero.

We expect new license issuance data to increase the model's ability to predict firearms simply because first-time purchasers require a license before legally purchasing a gun. License renewal data may improve the model's ability to predict firearm sales because a license renewal may correspond to the licensee's contemporaneous wish to legally acquire a new firearm.

*B. Estimation*

We estimate Equation (1) jointly for all four firearm types using the Seemingly Unrelated Regressions framework of Zellner (1962). SUR estimation improved model fit statistics slightly compared to single-equation OLS. R-square statistics were 0.74 for Shotgun, 0.77 for Non-TAW Rifles, 0.79 for TAW Rifles, and 0.82 for Handguns.

*C. Confidence Intervals for Predictions*

We make predictions and construct associated confidence intervals using the forward bootstrap with fitted residuals algorithm of Pan and Politis (2016). We draw from the asymptotic distributions of the parameter estimates and bootstrap fitted residuals to generate pseudo-series for each draw and iteratively predict post-policy outcomes. The approach accounts for both estimation error in the parameters and innovation errors without imposing distributional assumptions on the errors.

*D. Model Predictive Performance*

We evaluate the model's predictive performance shortly before the time of the Enforcement Notice. We re-estimated the model using data up until July 9, 2016; in other words, we held out the final 10 days of data before the EN (July 10—19, 2016). Figure 7 compares the model's sales predictions to actual sales observed during the 10-day holdout period.

Although this 10-day window occurred before the EN, it also occurred 28-37 days after the highly publicized Orlando mass shooting on June 12. Some of the purchases during this time were likely motivated by that anomalous event. Therefore, this is a challenging test of the model's predictive performance.

The model underpredicted cumulative firearm sales during this 10-day period by 11.6%. Mean daily prediction errors were –9% for Handguns, –12% for Shotguns, –14% for Non-TAW Rifles, and –21% for TAW Rifles. Daily prediction errors were greatest on July 10 and 11, at 43% and 46%; their range in the following eight days spanned –17% on July 16 to +24% on July 13. Observed sales generally fell within predictions' confidence bounds, but those bounds are fairly wide due to their accounting for both estimation error and innovation error.

Overall, we conclude that the model is able to predict firearm sales reasonably well around the time of the EN. Its mild underestimation of firearm sales may indicate that post-EN firearm sales predictions are conservative.

## VI. CAUSAL EVIDENCE

We compare observed firearm sales, by type, to the model's predictions of what counterfactual sales would have been in the absence of the EN. Given the large, five-day spike in rifle sales on the date of the EN announcement, we distinguish between an *immediate* effect occurring on the EN announcement date and the four days following, and a *short-run* effect that occurred from 5-25 days after the EN.[26]

Figure 8 graphs actual and predicted sales for the immediate effect. In the five days after the EN, consumers purchased 1,089 (+616%) more TAW Rifles than predicted, and 1,528 (+433%) more Non-TAW Rifles than predicted. The large majority of the surplus sales occurred on the date of the EN announcement. Both increases are statistically significant at the 95% confidence level. We speculate that the large, immediate effect of the EN on Non-TAW Rifle sales may reflect confusion in the marketplace and the difficulty of distinguishing banned semiautomatic assault rifles from visually similar rifles that are not banned.

The EN had smaller immediate impacts on Handgun and Shotgun sales. Handgun sales increased by 290 (+29%) in the five-day period after the EN. Shotgun sales increased by 77 (+63%) after the EN. Neither increase was statistically significant at the 95% confidence level.

How much of the sales spike represents "last-chance" purchasing among firearms enthusiasts or collectors vs. "commercially-motivated" purchases by professional firearms traders who anticipated appreciation on the used firearms market? The 5-day sales spike primarily consists of regular retail buyers: single-transaction purchasers bought 59.6% of Rifles sold during the spike, and two-transaction purchasers bought an additional 22.1%. However, the other end of the distribution may show some commercial motivations as well: 30 buyers bought between 5-15 Rifles each, and a single buyer purchased 27 Rifles, collectively accounting for 8.3% all Rifle purchases during the spike. As a point of reference, single-transaction purchasers accounted for 86% of all Rifle sales, and buyers with 5 or more purchases accounted for just 3.3% of Rifle purchases, in the 5-day period immediately preceding the spike.[27]

---

[26] We focus on a 25-day window out of concern that post-EN firearm license applications might affect the license issuances instrument. Similar conclusions are reached if we focus on the broader 5-35-day window instead.

Figure 9 graphs the short run effects from 5-25 days after the EN, by type of firearm. Firearm sales observations track model predictions closely. Cumulative observed TAW Rifle sales were 49 below the predicted level (-9%). Non-TAW Rifle sales were 194 below cumulative predictions (-15%). Neither cumulative difference is statistically significant.

Cumulative observed short-run Handgun sales were 244 (+6%) more than the model predicted. Cumulative observed short-run Shotgun sales were 50 (+8%) more than the cumulative prediction. Neither difference is statistically significant.

In summary, the EN caused large immediate increases in both TAW and Non-TAW Rifle sales, followed by short-run decreases of 9% and 15%, respectively. If the 9% reduction in short-run TAW Rifle sales were permanent, it would have taken 67 weeks for the negative short-run effect of the EN to counterbalance the positive immediate impact of the EN. If the 15% reduction in short-run Non-TAW Rifle sales were permanent, it would have taken 24 weeks for the negative short-run effect of the EN to counteract the positive immediate impact of the EN.

## VII. LONGER-TERM DATA DESCRIPTIVES

There is no control group or instrument available to separately identify longer-term causal effects of the EN from other factors (e.g., the 2016 election, mass shootings, or unknown drivers of firearm supply and demand). Therefore, we do not attempt to estimate long-run causal effects of the EN. Still, analysis of long-run data descriptives may be relevant and informative.

---

[27] We also checked whether patterns of retailer sales differed immediately before vs. during the spike. We found that the top 10 retailers accounted for 34.7% of Rifle sales during the 5-day spike, slightly less than the top 10 sellers' 39.1% collective share in the 5-day period preceding the spike.

*A. Firearm Sales in 2017*

Figure 10 graphs weekly firearm sales, by type and week of the calendar year, with separate lines for each year from 2014 until 2017. The annual overlays show limited seasonality in firearm purchases. The spikes in TAW Rifle and Non-TAW Rifle sales in mid-2016 remain prominent. The smaller spike in June 2016 corresponded to the mass shooting in Orlando on June 12.

Figure 10 also shows that consumers purchased more firearms—of all types—in October 2016 than in October 2014 or 2015. We think that was related to the U.S. presidential election season, as the two previous presidential election seasons also coincided with large increases in firearm sales. In the final few weeks of 2016, it appears that TAW Rifle sales fell below 2014 and 2015 levels, although sales of the three other types of firearm did not fall below 2014 or 2015 levels in the same period. This was a rare divergence in sales trends between TAW rifles and Non-TAW Rifles.

In 2017, TAW Rifle sales hovered near the bottom end of the range of 2014-2016 levels for nearly the entire year. The graph shows a downward shift in the level of TAW Rifle sales with fewer large spikes, but no clear downward trend within 2017. For example, the average change in monthly TAW Rifle sales between 2015 and 2017 was -28%. The average change in monthly TAW Rifle sales in the first half of 2016 vs. the first half of 2017 was similar at -30%. No similar downward shift is observed in 2017 weekly sales vs. previous years for Non-TAW Rifles, Shotguns or Handguns.

The fact that TAW Rifle sales were 28-30% lower in 2017 than in comparable previous periods, along with the absence of 2017 sales decreases of other types of weapon, suggests that the EN partially succeeded in decreasing assault rifle sales in the long run. However, recall that the TAW Rifle classifier is conservative and likely to undercount the proportion of assault rifles sold in the marketplace. If the TAW Rifles distinction accurately measures sales of AR-style rifles, this pattern also suggests frequent noncompliance with the EN by some retailers and consumers. For example, if the EN applied to all AR-style rifles, and if the measure of TAW Rifles perfectly measures AR-style rifles, perhaps the short-run decrease in TAW Rifles should have been closer to 100%. The substantial gap between a 28-30% observed drop and a hypothetical 100% drop suggests that many illegal firearm sales occurred in 2017.

We can speculate about explanations for the widespread noncompliance with the Enforcement Notice. One is civil disobedience: it is possible that many firearm sellers and purchasers knew about the EN and deliberately disregarded it. Another is ignorance. Some market participants simply may not know what actions are legal or not legal; for example, they may not know how to properly apply the Interchangeability Test or Similarity Test. It is relevant to consider that the data we observe was reported by sellers to the state through legal channels; some dealers might prefer to avoid knowingly informing the state about illegal weapon sales. Third, as the AGO made clear, the AGO relied on dealers' voluntary compliance with the law. Such compliance may have been difficult or unprofitable. Finally, we have found no evidence that assault weapons ban enforcement efforts increased concurrently with the EN. In fact, the AGO publicly requested voluntary compliance, and its communications suggested that some dealers were known to be out of compliance. It may be that increased enforcement is required to gain a higher level of compliance in the marketplace.

*B. Sales to Newly-observed Purchasers*

How much of the 2016 sales spike was caused by new gun purchasers as opposed to repeat purchasers or collectors? The data contain anonymous purchaser identifiers, so we can partially address this question. We identified the first week each anonymous purchaser identifier appeared in the data, and then counted purchases by that identifier in its first observed week as sales to "newly-observed purchasers."[28]

Figure 11 graphs weekly sales to newly-observed purchasers, by firearm type and week of year, with separate lines for each year from 2014 until 2017. Newly-observed purchasers accounted for about 8% of the EN-related sales spike in TAW Rifles, and about 7% of the EN-related sales spike in Non-TAW Rifles. Handguns and Shotguns did not show concurrent spikes. Sales of TAW rifles to newly-observed purchasers remained strictly positive after the EN, though 2017 sales were observably lower than in previous years.

*C. Retailer Changes in TAW Rifle Sales*

Although TAW Rifle sales fell in 2017 compared to earlier years, many TAW Rifle sales continued to occur. How were those sales distributed across retailers and geographies? The answers might offer novel evidence about how firearm sellers voluntarily comply with firearm policies. To answer these questions, we constructed a subsample of 59 retailers which collectively accounted for 90% of all TAW Rifle sales in 2015. Then, for each retailer, we examined how TAW Rifle and Non-TAW Rifle sales changed from 2015 to 2017.[29]

There are some interesting patterns. The largest TAW Rifle seller in 2015, Cabela's Wholesale Inc., in Hudson, Mass., saw its TAW Rifle sales fall by 91% from 2015 to 2017, from 1,341 units down to 115 units. Meanwhile, the same retailer's Non-TAW Rifle sales rose from

---

[28] Some of the newly-observed purchasers may have first purchased weapons in other states or prior to 2006.
[29] We compare 2015 and 2017 because both are non-election years without major MA firearm policy changes.

81 units in 2015 to 1,959 units in 2017. These changes suggest that the retailer intentionally stocked different firearms in order to comply with the EN. Similarly, the second-largest TAW Rifle seller in 2015, Bass Pro Outdoor Worldwide LLC, in Foxborough, Mass., reduced its TAW Rifle sales by 84%, from 624 in 2015 to 99 in 2017. It simultaneously increased its Non-TAW Rifle sales from 192 in 2015 up to 544 in 2017.

Some retailers' sales data show different patterns. Shooting Supply LLC, in Westport, MA, increased its TAW Rifle sales 85% from 104 in 2015 up to 192 in 2017. Its Non-TAW Rifle sales increased by a similar 86%, from 124 in 2015 up to 231 in 2017. Because the changes in TAW Rifle sales and Non-TAW Rifle sales were similar, it seems most likely that the seller either was not aware of the EN or did not adjust its inventory in response to the EN. It is notable that we did not find any patterns of retailer sales changes that suggested that individual sellers dramatically increased TAW Rifle sales without similar increases in Non-TAW Rifle sales.

Figure 12 provides a more comprehensive look at how TAW Rifle sales changed across retailers. For each of the 59 largest TAW Rifle sellers in 2015, and for each firearm type, we calculated the ratio of 2017 sales to 2015 sales. Figure 12 shows the histograms of retailer sales ratios across firearm types. Most sellers reduced their TAW Rifle sales considerably while simultaneously increasing their Non-TAW Rifle sales considerably, suggesting substitution between these two classes of weapon. Handgun and Shotgun sales ratios generally show smaller increases than Non-TAW Rifle sales, which may be consistent with the short-run null effects of the EN on Handgun sales and Shotgun sales.

Finally, we sought to understand how changes in TAW Rifle sales correspond to geographic characteristics, in an effort to understand how EN compliance varied across regions. For each of the 59 retailers mentioned above, we found their zip codes in the Federal Firearm

Licensee directory provided by the U.S. Bureau of Alcohol, Tobacco and Firearms.[30] We then linked retailer sales ratios to zip code characteristics from the U.S. Census.

Figure 13 relates retailers' 2017/2015 TAW Rifle sales ratios to retailer zip codes' median household income. 2017 TAW Rifle sales fell less in high-income zip codes, whereas sales ratios for other firearm types are nearly unrelated to zip codes' income levels. Figure 14 shows that TAW Rifle sales ratios fell less in zip codes with higher proportions of white males, a pattern which is stronger than that observed for other firearm types. Although these results are merely suggestive, they may provide some basis for deployment of enforcement resources across areas. They also offer suggestive evidence about which areas voluntarily complied with the EN, and they help to illustrate the value of state governments collecting and disseminating firearm sales data.

**VIII. DISCUSSION**

We offer rare causal evidence of how firearm policy changed firearm sales. We introduce and evaluate an approach to classify rifles as assault weapons or not using an automated image classifier. We estimate a model of firearm sales up until the point of the Enforcement Notice, and then measure causal effects of the EN by comparing the model's predictions to observed sales data in the 25 days after the announcement. We found that the Enforcement Notice caused large, immediate increases of sales of TAW Rifles and Non-TAW Rifles, followed by smaller short-run decreases. Shotgun and Handgun sales were not significantly changed after the EN. Data descriptives show that TAW Rifle sales were about 28-30% lower in 2017 than in comparable previous years, suggesting that the EN reduced legal sales of semiautomatic assault rifles, and also that many banned assault weapons continued to be sold.

---
[30] https://www.atf.gov/firearms/listing-federal-firearms-licensees

Several aspects of the results might be relevant to policy makers. First, it is apparent that firearm sales may react rapidly to firearm policy changes. The rapid reaction may reflect forward buying by consumers. It may also be that the firearm industry amplifies policy-related news in order to stimulate immediate sales. Policy makers may wish to anticipate market reactions and design their public communication strategies accordingly.

Second, we have found evidence suggesting that the EN was partially effective in reducing assault weapon sales. This effect occurred without any accompanying revision of state law. The result may be relevant to assault weapons ban enforcement and interpretation in other jurisdictions.

Third, we found that TAW Rifles and non-TAW Rifles both showed large increases at the time of the EN. The large, immediate increase in Non-TAW Rifle sales suggests possible confusion in the market about which weapons were banned or not banned, as well as likely reflecting the imperfect ability of the automated classifier to distinguish between banned and non-banned Rifles. The first interpretation is further buttressed by arguments in firearm retailer lawsuits filed after the EN (see, e.g., Petrishen 2019), in which retailers claimed that the AGO was vague in its communications about which specific firearm products were legally banned or not. Banning weapons based on feature tests has been criticized as "porous" (Donohue 2012), and has led weapon manufacturers to skirt legal bans by modifying firearm features, such as using replaceable magazines in place of detachable magazines (White 2019). If the policy goal of banning assault weapons is to is to restrict or ban weapons whose capabilities facilitate mass shootings—such as rapid firing and rapid reloading—it may be advisable to provide greater clarity to market participants about which weapons are banned and which are not banned. It also may be advisable to consider writing regulations around weapon capabilities, rather than weapon features. These suggestions are informed speculation; they do not follow directly from the empirical analysis.

Finally, this research has implications for state and local governments regarding firearm data collection and provision. For example, it might be advisable for Massachusetts state authorities to modify their data collection procedure to require firearm dealers to enter the specific features of the individual weapons they sell. Doing so could facilitate automated detection of the two-feature test or other measures of compliance with state law. It may also be advisable to post anonymous firearm sales data to the public to facilitate third-party detection of important trends. More and better scientific analyses will become possible if more jurisdictions collect and disseminate privacy-compliant firearm sales data.

The current study has some important limitations. The data come exclusively from Massachusetts, a state that likely has some of the strictest firearm policies and lowest firearm ownership rates in the nation. The Massachusetts results are interesting for their own sake but they may not be representative of other jurisdictions. Second, the sales data were recorded by firearm dealers. We have limited ability to audit or verify the data. Third, the available data have limited ability to explain the mechanisms underlying the main causal effects. For example, we do not observe intermediate drivers of market outcomes such as prices, EN-related publicity (e.g., emails from firearm retailers to consumers), consumer search or store traffic, so the data provide limited means to disentangle forward buying from other explanations for the sales spike and subsequent decrease. Finally, we have not connected the legal purchase data to other possible outcomes, such as the used guns market, the illegal firearm market, or crimes such as homicide or aggravated assault. We focus on legal firearm acquisitions as that is the behavior the assault weapon ban and EN are expressly designed to limit, and because guns used in crimes are typically traded on the used firearm market before criminal use, and seldom used by their initial purchasers. We think it is unlikely that legal firearm purchases by licensed gun owners who passed background checks would sharply increase crimes in the short run, but it remains possible that the legal sales spike may have expanded firearm supply in other markets.

In summary, this paper seeks to contribute to the growing body of empirical knowledge about firearm regulation by evaluating a new technique to classify assault weapons, by analyzing the immediate impacts of the EN on firearm sales, and by providing evidence suggesting partial compliance with the EN in the longer run. We are hopeful that the body of empirical firearm research will continue to grow. If evidence-based firearm policy is a desirable goal, we will need more and better firearm science to help inform policymakers of policy effects.

Disclosures:

Authors have no funding or conflicts of interest to report. We undertook this research with the goal of providing empirical facts to help inform policy makers. As IOM and NRC (2013) put it, "In the absence of research, [firearm] policy makers will be left to debate controversial policies without scientifically sound evidence about their potential effects." We will post final data and estimation code online.

Roth, J. A., & Koper, C. S. (1999). Impacts of the 1994 Assault Weapons Ban: 1994-96. National Institute of Justice: Research in Brief, March 1999, 1-12.

Smart, R., Morral, A. R., Smucker, S., Cherney, S., Schell, T. L., Peterson, S., Ahluwalia, S. C., Cefalu, M., Xenakis, L., Ramchand, R., & Gresenz, C. R. (2020). The science of gun policy: A critical synthesis of research evidence on the effects of gun policies in the United States, Second Edition. Santa Monica, CA: RAND Corporation. https://www.rand.org/pubs/research_reports/RR2088-1.html.

Studdert, D. M., Zhang, Y., Rodden, J. A., Hyndman, R. J., & Wintemute, G. J. (2017). Handgun acquisitions in California after two mass shootings. Annals of Internal Medicine, 166, 698-706.

White, J. (2019). When Lawmakers Try to Ban Assault Weapons, Gunmakers Adapt. New York Times, July 31. https://www.nytimes.com/interactive/2019/07/31/us/assault-weapons-ban.html, accessed January 2021.

Zellner, A. (1962). An efficient method of estimating seemingly unrelated regressions and tests for aggregation bias. Journal of the American Statistical Association, 57, 348-368.

Zitzewitz, E. (2012). Forensic economics. Journal of Economic Literature, 50, 731-769.

**Table 1. Time Effect Specifications and 10-Fold Cross-Validation Prediction Errors**

| | | | | | | |
|---|---|---|---|---|---|---|
| Day-of-Week F.E. | x | x | x | x | x | x |
| Holiday F.E. | x | x | x | x | x | x |
| Day-of-Year F.E. | | x | | x | | x |
| Week-of-Year F.E. | x | | x | | x | |
| Linear Trend | | | x | x | x | x |
| Quadratic Trend | | | | | x | x |

*Prediction Errors in 10-fold Cross-Validation*

**Handguns**
| | | | | | | |
|---|---|---|---|---|---|---|
| Root Mean Sq. Prediction Error | 0.473 | 0.490 | 0.475 | 0.491 | **0.471** | 0.489 |
| Mean Abs. Prediction Error | 0.279 | 0.295 | 0.279 | 0.297 | **0.277** | 0.294 |

**TAW Rifles**
| | | | | | | |
|---|---|---|---|---|---|---|
| Root Mean Sq. Prediction Error | 0.508 | 0.531 | 0.507 | 0.528 | **0.505** | 0.525 |
| Mean Abs. Prediction Error | 0.370 | 0.388 | 0.369 | 0.388 | **0.366** | 0.383 |

**Non-TAW Rifles**
| | | | | | | |
|---|---|---|---|---|---|---|
| Root Mean Sq. Prediction Error | 0.470 | 0.488 | 0.469 | 0.483 | **0.465** | 0.479 |
| Mean Abs. Prediction Error | 0.323 | 0.340 | 0.322 | 0.336 | **0.320** | 0.334 |

**Shotguns**
| | | | | | | |
|---|---|---|---|---|---|---|
| Root Mean Sq. Prediction Error | 0.474 | 0.545 | 0.473 | 0.516 | **0.467** | 0.510 |
| Mean Abs. Prediction Error | 0.335 | 0.403 | 0.334 | 0.377 | **0.328** | 0.372 |

**Figure 1. Rifle sales spiked on the date of the EN announcement (2016-07-20), but Handgun and Shotgun sales did not change much.**

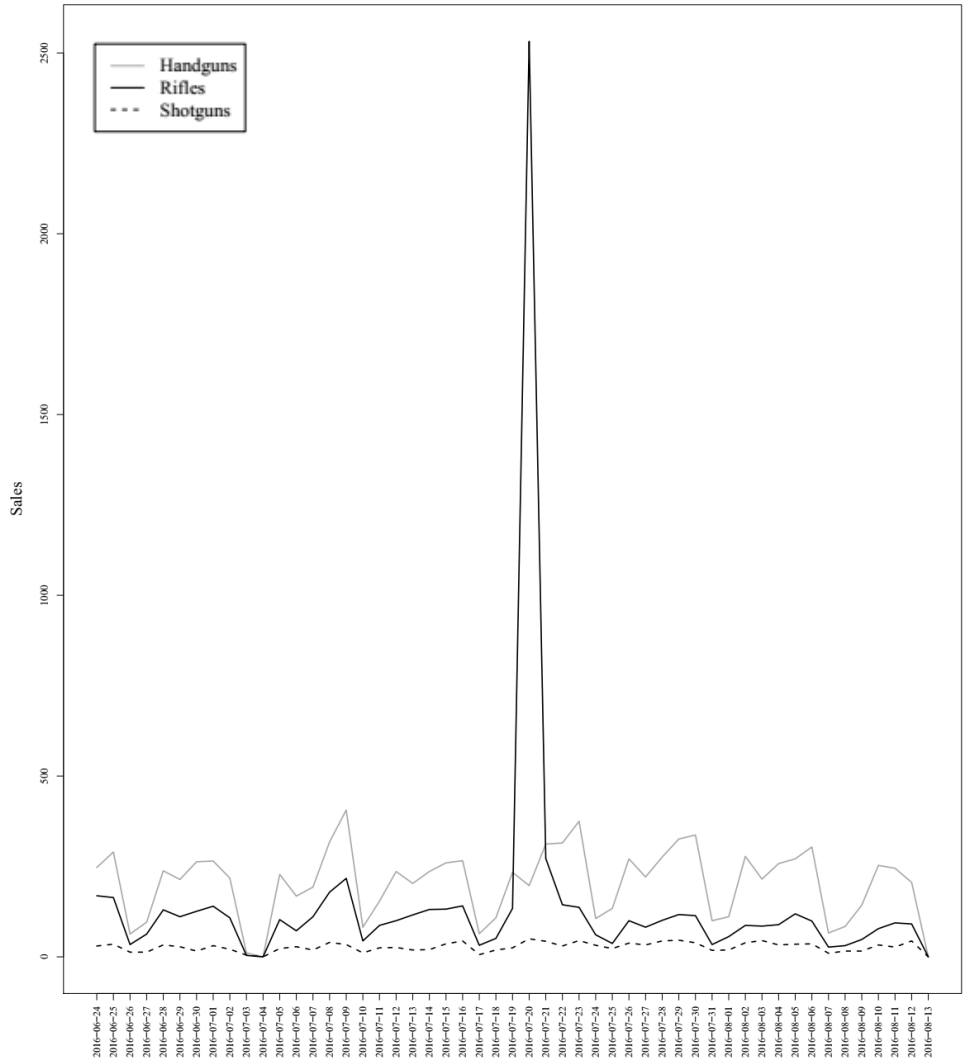

**Figure 2. Sample images for most frequent Rifle make/model combinations**

1. Ruger 10/22: 1,636 sales; Non-TAW

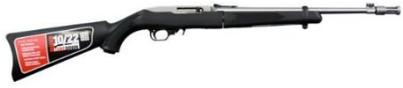

2. Ruger AR-556: 672 sales; TAW

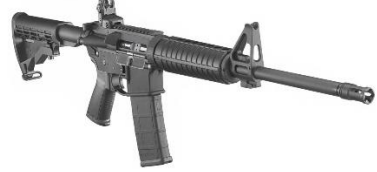

3. Windham Weaponry WW-15: 476 sales; TAW

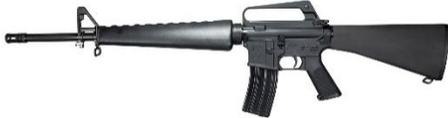

4. Ruger Mini 14: 462 sales; TAW

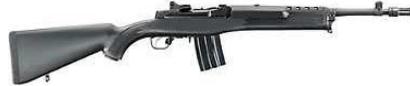

5. Remington 700: 426 sales; Non-TAW

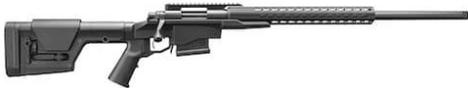

6. Ruger American: 395 sales; Non-TAW

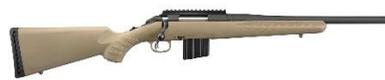

**Figure 3. Confusion Matrices to assess Google Cloud Vision classifier**

*Measure of Truth: Median RA classification (N=98 Rifles)*

| weighted equally | Google Vision: Not a TAW Rifle | Google Vision: TAW Rifle | |
|---|---|---|---|
| RAs: Not an Assault Rifle | True Negative 36 | False Positive 3 | 39 |
| RAs: Assault Rifle | False Negative 12 | True Positive 47 | 59 |
| | 48 | 50 | |

| weighted by sales | Google Vision: Not a TAW Rifle | Google Vision: TAW Rifle | |
|---|---|---|---|
| RAs: Not an Assault Rifle | True Negative 28.3 | False Positive 3.6 | 31.9 |
| RAs: Assault Rifle | False Negative 20.0 | True Positive 46.1 | 66.1 |
| | 48.3 | 49.7 | |

*Measure of Truth: Unanimous RA classification (N=69 Rifles; dropping 29 Rifles for which RAs were not unanimous in assault weapon classification)*

| weighted equally | Google Vision: Not a TAW Rifle | Google Vision: TAW Rifle | |
|---|---|---|---|
| RAs: Not an Assault Rifle | True Negative 27 | False Positive 0 | 27 |
| RAs: Assault Rifle | False Negative 4 | True Positive 38 | 42 |
| | 31 | 38 | |

| weighted by sales | Google Vision: Not a TAW Rifle | Google Vision: TAW Rifle | |
|---|---|---|---|
| RAs: Not an Assault Rifle | True Negative 21.5 | False Positive 0 | 31.9 |
| RAs: Assault Rifle | False Negative 4.2 | True Positive 43.2 | 66.1 |
| | 25.7 | 43.2 | |

**Figure 4. Annual Firearm sales by Type**

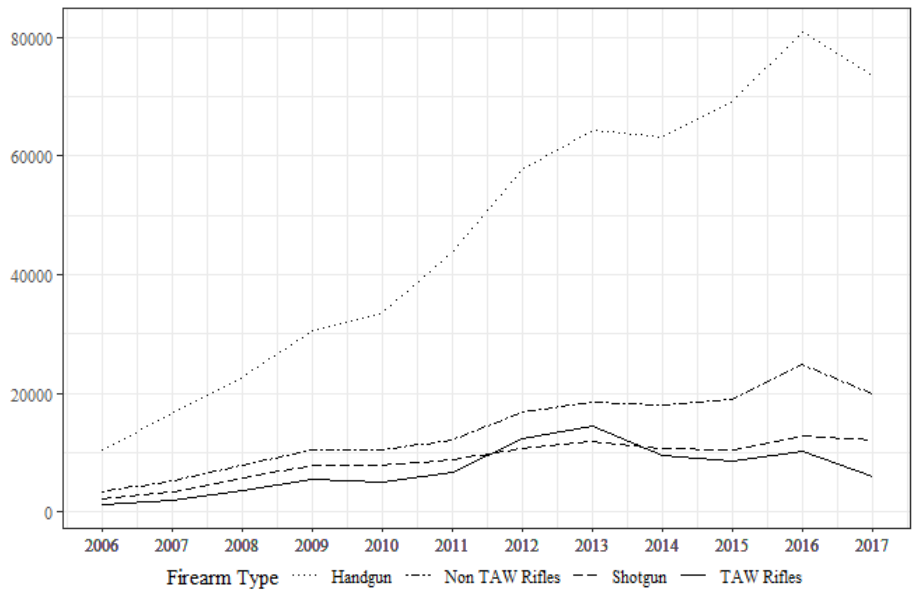

**Figure 5. Annual License Issuances by Type**

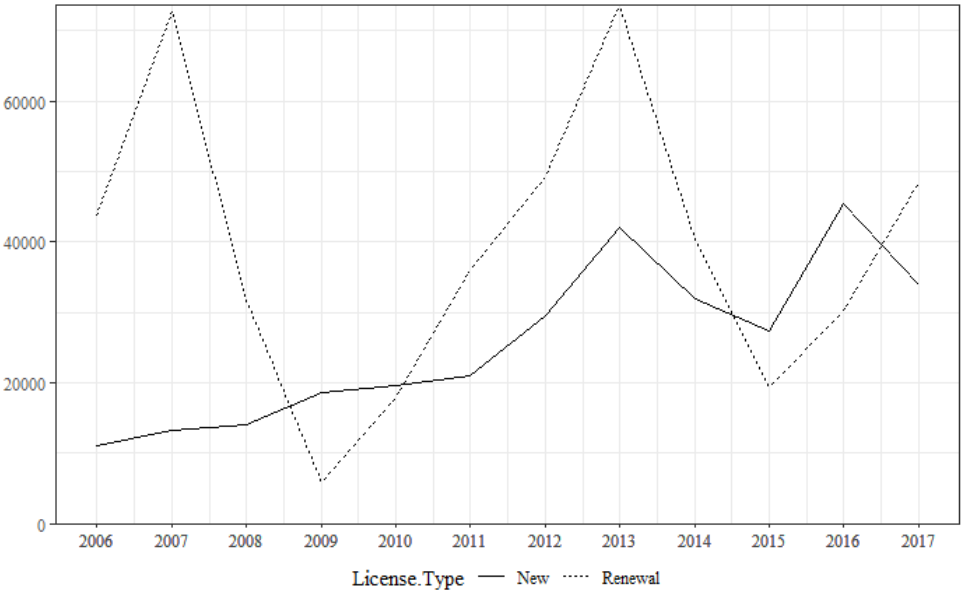

**Figure 6. Daily License Issuances Around the EN Date**

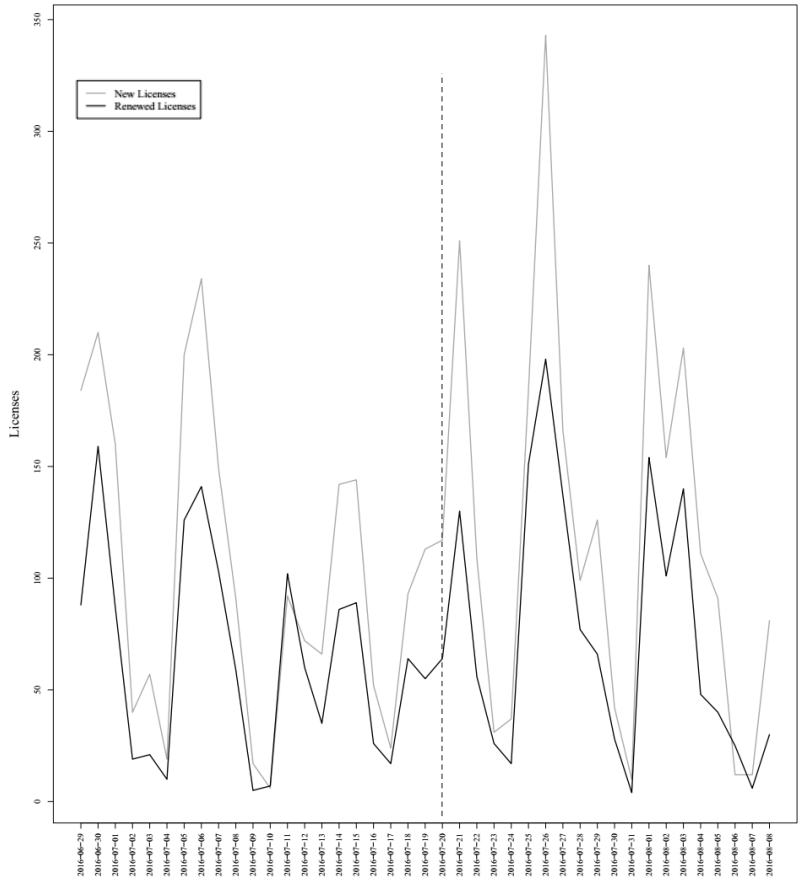

**Figure 7. Predicted Sales and Observed Sales before the Enforcement Notice**

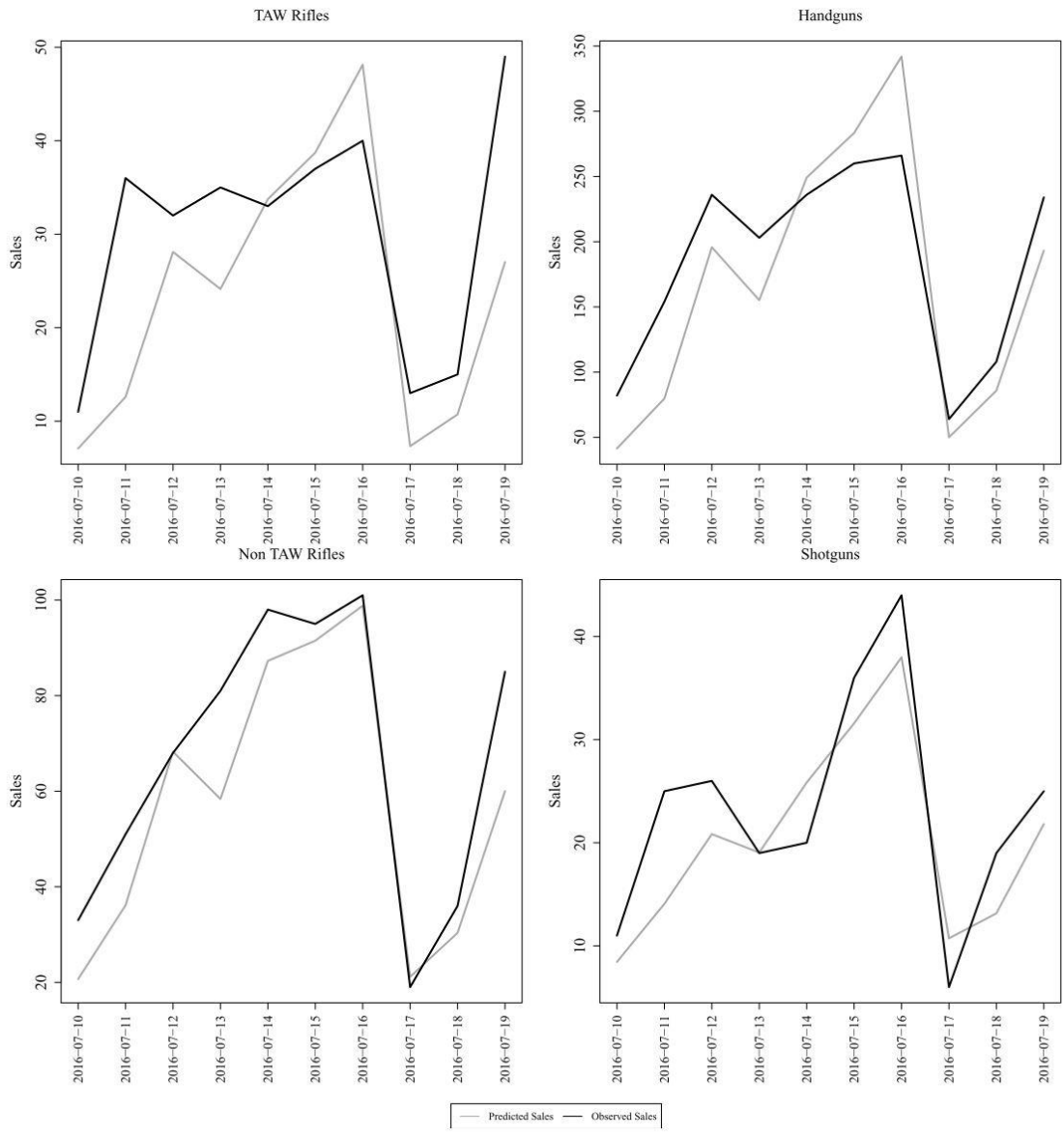

**Figure 8. Immediate effect of EN on Firearm Sales, by Type**

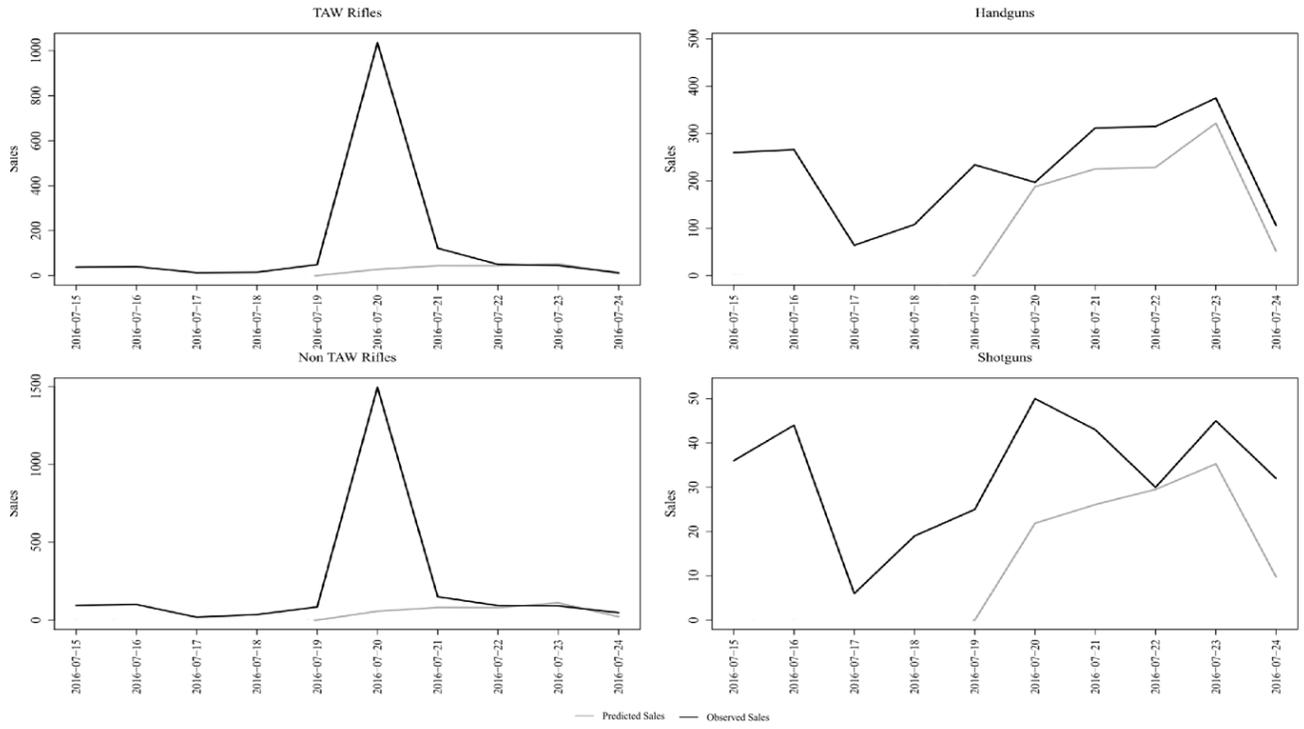

**Figure 9. Short-run effects of EN on Firearm Sales, by Type**

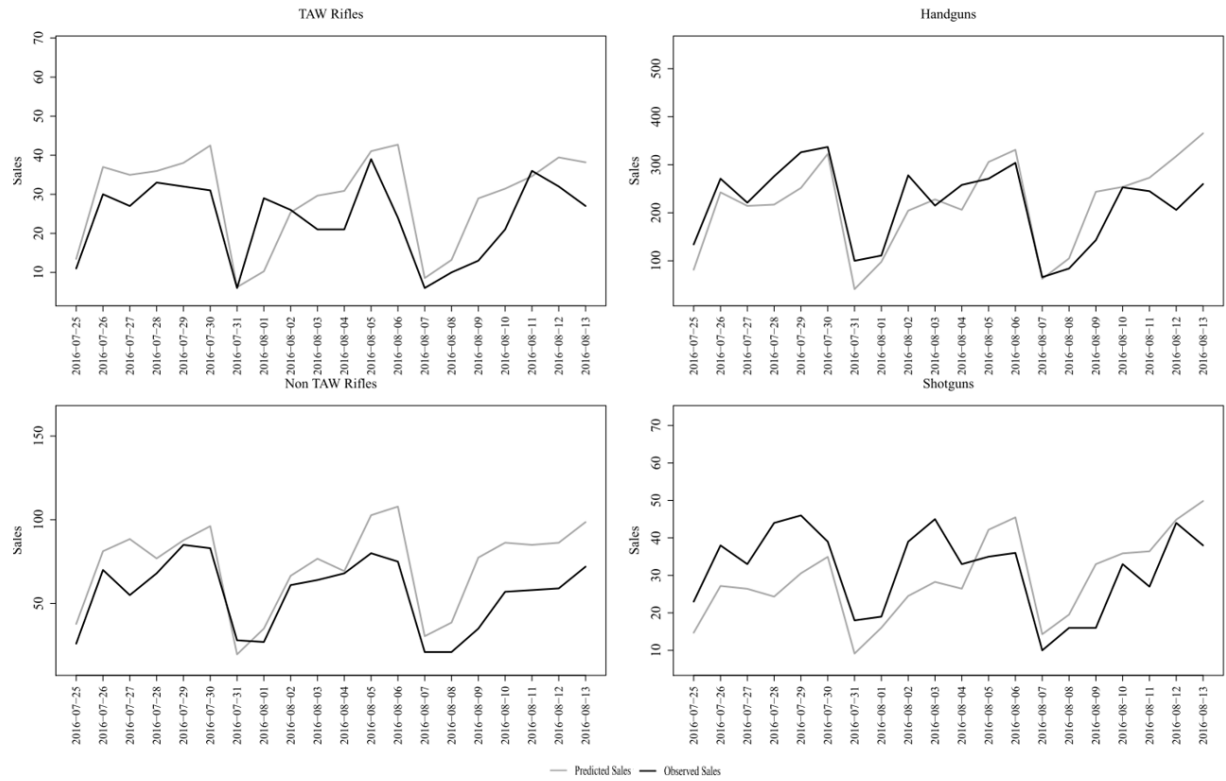

**Figure 10. Firearm sales in 2014-2017, by type and by week of year**

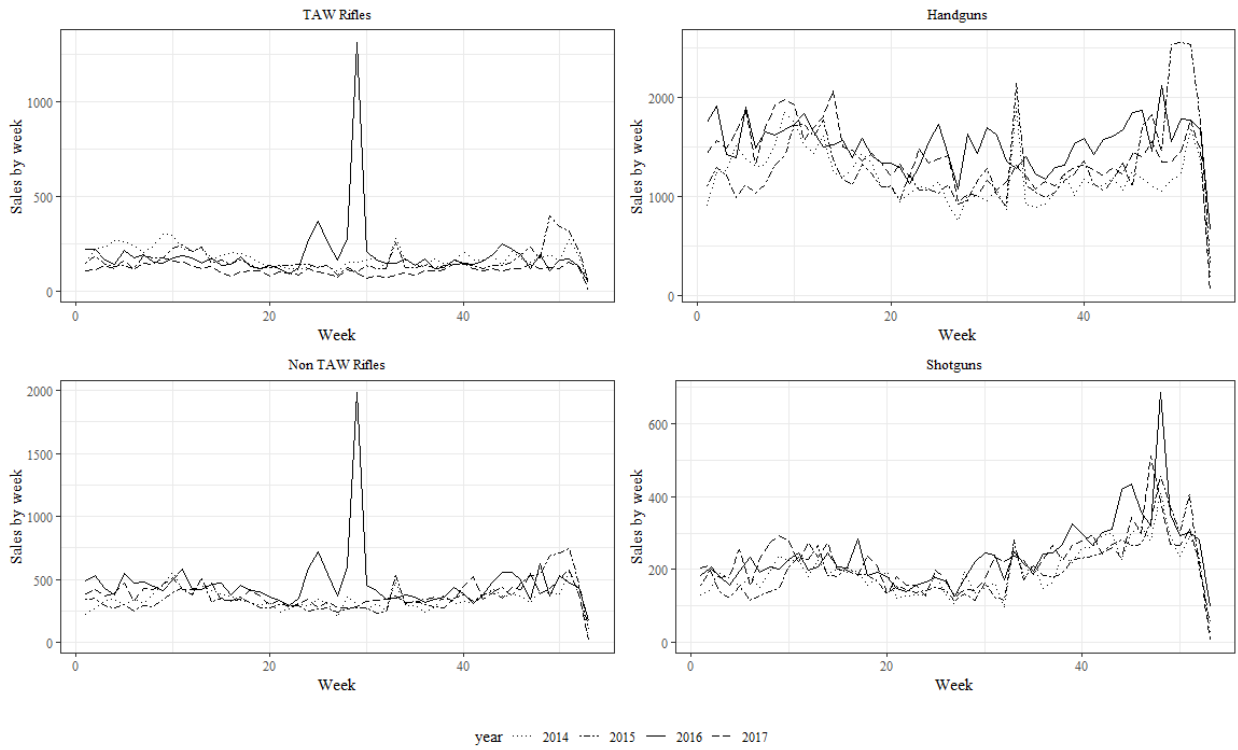

**Figure 11. Firearm Sales to Newly-observed Purchasers in 2014-2017, by Type and Week of Year**

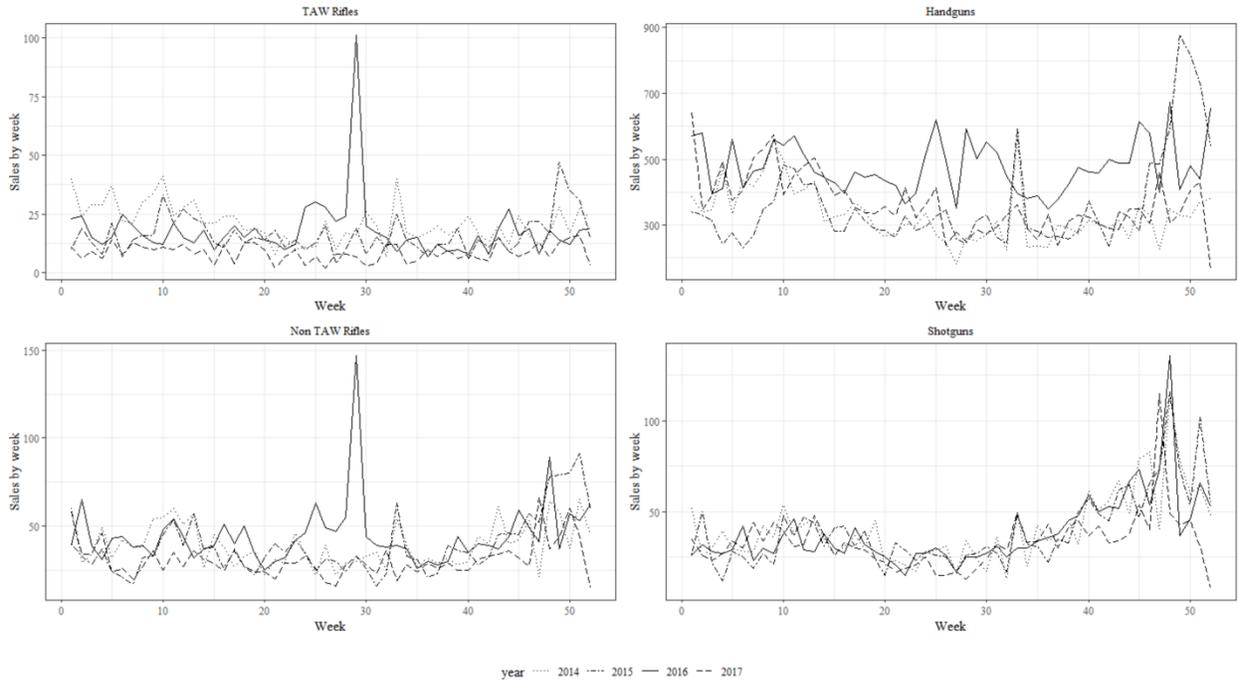

# Figure 12. Histograms of Retailer Sales Ratios

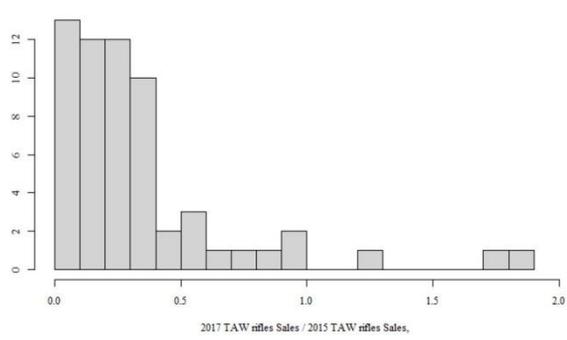
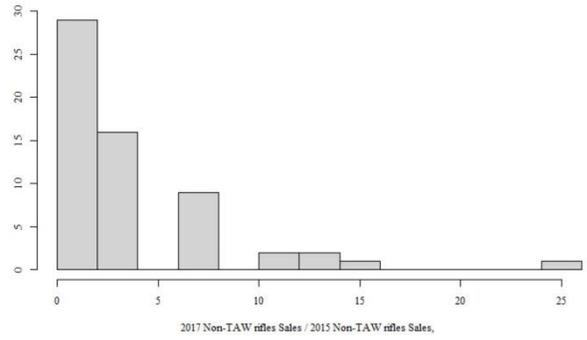
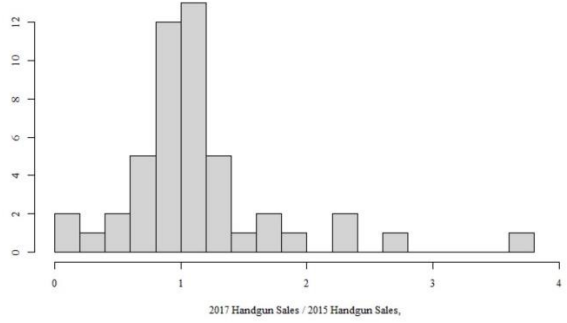
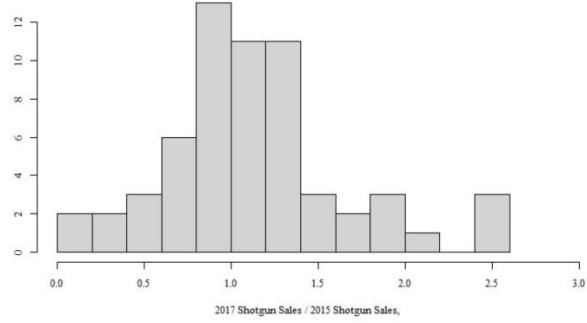

**Figure 13. Retailer Sales Ratios and Household Incomes by Zip Code**

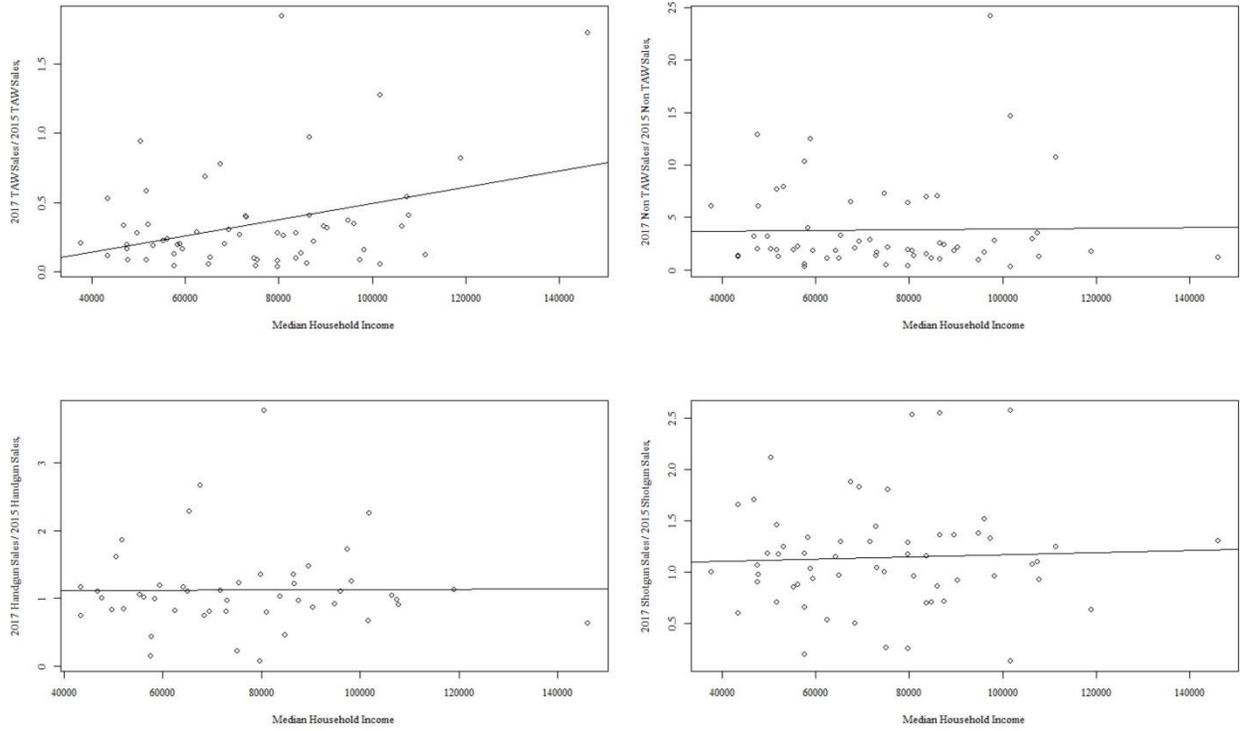

**Figure 14. Retailer Sales Ratios and Demographics by Zip Code**

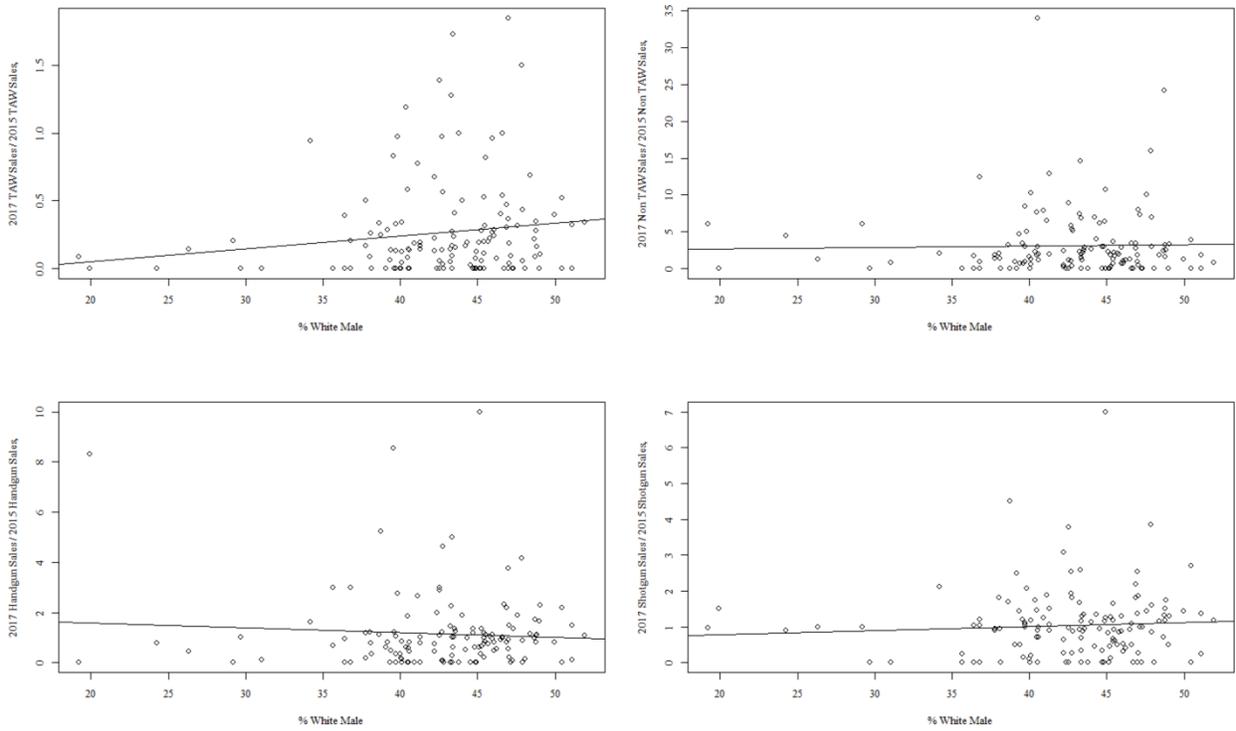